\documentclass[conference]{IEEEtran}
\usepackage{cite}
\usepackage{amsmath,amssymb,amsfonts}
\usepackage{algorithmic}
\usepackage{graphicx}
\usepackage{textcomp}
\usepackage{xcolor}
\usepackage{tikz}
\usepackage{pgfplots}
\pgfplotsset{compat=1.18}
\usepackage{subcaption}
\usepackage{url}
\usepackage{subcaption}
\begin{document}

\title{From Individual Prompts to Collective Intelligence: Mainstreaming Generative AI in the Classroom}

\author{
\IEEEauthorblockN{
Junaid Qadir and Muhammad Salman Khan
}
\IEEEauthorblockA{
College of Engineering, Qatar University, Doha, Qatar\\
Emails: jqadir@qu.edu.qa, salman@qu.edu.qa
}
}

\maketitle

\begin{abstract}
Engineering classrooms are increasingly experimenting with generative AI (GenAI), but most uses remain confined to individual prompting and isolated assistance. This narrow framing risks reinforcing equity gaps and only rewarding the already privileged or motivated students. We argue instead for a shift toward collective intelligence \textit{(CI)-focused pedagogy}, where GenAI acts as a catalyst for peer-to-peer learning. We implemented Generative CI (GCI) activities in two undergraduate engineering courses, engaging 140 students through thinking routines---short, repeatable scaffolds developed by Harvard Project Zero to make thinking visible and support collaborative sense-making. Using routines such as Question Sorts and Peel the Fruit, combined with strategic AI consultation, we enabled students to externalize their reasoning, compare interpretations, and iteratively refine ideas. Our dual-pronged approach synthesizes literature from learning sciences, CI, embodied cognition, and philosophy of technology, while also empirically learning through student surveys and engagement observations. Results demonstrate that students value the combination of human collaboration with strategic AI support, recognizing risks of over-reliance while appreciating AI's role in expanding perspectives. Students identified that group work fosters deeper understanding and creative problem-solving than AI alone, with the timing of AI consultation significantly affecting learning outcomes. We offer practical implementation pathways for mainstreaming CI-focused pedagogy that cultivates deeper engagement, resilient problem-solving, and shared ownership of knowledge. 
\end{abstract}


\section{Introduction}
\label{sec:introduction}

Engineering education stands at a critical juncture. Generative AI (GenAI) tools are now ubiquitous in classrooms \cite{qadir2023engineering}, yet their use remains largely individualistic---students consulting AI privately to generate answers, code, or explanations for personal efficiency. This pattern runs counter to how engineering expertise actually develops: through social interaction, collaborative reasoning, apprenticeship-like engagement with authentic problems, and productive struggle \cite{vygotsky1978mind,bonwell1991active}. Research on situated learning further underscores the role of situated judgment---the ability to interpret and act within real contexts---which emerges through participation in authentic practice rather than in individualistic tasks \cite{johri2011situated}.

The consequences of this misalignment are increasingly evident: students become dependent on AI-generated outputs and show weakened critical thinking, analytical reasoning, and independent problem-solving, leading to noticeable drops in performance when AI support is removed \cite{qadir2025generative}. Academic integrity concerns rise as it becomes increasingly difficult to distinguish authentic student work from AI-generated content. Students begin to conflate AI's abilities with their own and display confidence that is not matched by competence \cite{qadir2024learning}, as polished AI outputs provide an ``illusion of learning.'' Faculty also report a breakdown in traditional mentor–mentee relationships as students replace office hours, peer discussion, and iterative feedback with instantaneous AI responses. 

Many of these challenges stem from implementation models that treat AI as an individual tool rather than a facilitator of collective intelligence (CI) \cite{malone2022handbook,casebourne2025using}. In a Generative Collective Intelligence (GCI) framework \cite{kehler2025amplifying}, GenAI functions as a connective layer linking human reasoning with computational capabilities to enable forms of collaborative problem-solving that neither humans nor AI could achieve alone. Similarly, Moldoveanu and Siemens' \cite{moldoveanu2025interactionalism} framework emphasizes ``\textit{interactional intelligence}'', noting that modern work is predominantly group-based and requires not only individual expertise but also the capacity to think, learn, and create in dialogue with human collaborators and AI agents.

\subsection{Motivation and Significance}

This paper addresses an urgent need in education: moving beyond individualistic AI use toward pedagogical approaches that leverage AI to amplify collective human intelligence while maintaining the social, embodied, and relational dimensions essential for authentic learning and human flourishing \cite{aaker2024humanity}. Three converging crises motivate this work:

\begin{enumerate}
\item \textit{Cognitive Debt Crisis.} Research from the MIT Media Lab \cite{kosmyna2025cognitive} provides neurocognitive evidence that students who relied on ChatGPT during writing tasks exhibited weaker neural connectivity and significantly poorer recall than peers who generated ideas independently. This pattern reflects a cognitive debt: immediate gains in ease and efficiency are borrowed at the expense of long-term memory, deep processing, and sense-making. 

\item \textit{Social Learning Crisis.} Engineering is inherently collaborative. Yet individualistic AI use isolates learners from the peer interaction, diverse perspectives, constructive disagreement, and collective meaning-making that characterize both effective pedagogy and professional engineering practice. When students consult AI privately rather than reasoning collectively, they lose opportunities for the socially-situated learning that develops both technical mastery and the communication, teamwork, and ethical reasoning essential for engineering careers.

\item \textit{Equity and Inclusion Crisis.} Individual AI usage rewards privileged and motivated students, who already possess strong metacognitive skills and have clear learning goals. Students lacking these advantages, or from underrepresented backgrounds, gain less while experiencing the same cognitive risks. CI approaches, by contrast, distribute cognitive load, leverage diverse perspectives, and provide scaffolding through peer support, creating more equitable learning environments.

\end{enumerate}

\textit{CI-focused pedagogy} directly addresses these crises by preparing students for a future in which AI is ubiquitous yet the very capacities required to guide, critique, and ethically steer its use---creativity, judgment, ethics, and collaboration---are the ones most vulnerable to erosion through overreliance on AI \cite{qadir2024educating}. Rather than training students to compete with machines that will outperform humans on many technical tasks, CI-oriented education strengthens what remains distinctly human: critical questioning, ethical reasoning, creative problem framing, empathetic collaboration, and the embodied, contextual judgment that AI fundamentally lacks \cite{aaker2024humanity}.

\subsection{Contributions}

Towards this end, this paper makes two contributions to engineering education scholarship and practice. 
 
\begin{itemize}
 
\item \textit{Theoretical Synthesis}---We integrate frameworks from learning sciences, CI theory, embodied cognition, and philosophy of technology to propose a pedagogical approach that prioritizes collective over individual intelligence in AI-augmented engineering classrooms.

\item \textit{Implementation and Exploratory Evidence}---We provide detailed activity designs combining Making Thinking Visible routines \cite{ritchhart2020power} with strategic AI consultation, and present preliminary findings from 140 students across two engineering courses documenting student perceptions of different learning modalities, timing effects of AI consultation, and experiences comparing individual AI use with collective approaches.
\end{itemize}

\subsection{Organization}

The remainder of the paper is organized as follows. Section \ref{sec:background} reviews key foundations from social learning theory and the limitations of individualistic AI use. Section \ref{sec:GCI} introduces GCI-related pedagogical theories alongside our implementation framework. Section \ref{sec:methodology} outlines our dual-pronged approach combining literature synthesis with classroom implementation. Section \ref{sec:results} reports findings from student surveys and observations. Section \ref{sec:discussion} interprets the results, discusses implications, notes limitations, and identifies future directions. Section \ref{sec:conclusion} synthesizes key insights and offers recommendations for advancing CI pedagogy in engineering education.

\section{Background}
\label{sec:background}

\subsection{The Social and Active Nature of Learning}

\subsubsection{Active Nature of Learning}
Complementing this, research on active learning demonstrates that students learn more effectively when they are cognitively and behaviorally engaged (e.g., using strategies such as discussion, problem-solving, and reflection) rather than as passive recipients of information \cite{bonwell1991active}. These insights underscore a central tension in individualistic AI use: while AI accelerates task completion, it can also bypass the very social and active processes through which expertise is built. Engineering expertise in particular emerges through collaborative problem-solving, peer explanation, and apprenticeship-like engagement with complex tasks.

\subsubsection{Embodied Dimensions of Learning}

Beyond information processing, learning involves embodied and relational dimensions that purely cognitive approaches cannot replicate. Phenomenological traditions---from Heidegger \cite{heidegger1962being} through Dreyfus \cite{dreyfus1992computers} and Varela \cite{varela1991embodied}, to their application in computing by Winograd and Flores \cite{winograd1986understanding}---demonstrate that cognition emerges from embodied engagement with environments, tools, and people. In engineering, expertise develops through the messy work of translating abstract principles into working systems; grappling with open-ended design problems; analyzing trade-offs, iterating through failures, and cultivating practical judgment. AI assists with information processing but cannot substitute for the embodied practice through which engineering competence actually forms. The relational and humanistic dimensions are equally essential \cite{aaker2024humanity}. 

\subsubsection{Relational Dimensions of Learning}

Learning occurs within relationships that provide not just knowledge transfer but identity formation (becoming an engineer), ethical development (understanding professional responsibilities), and social support for persistence through challenge \cite{schon1983reflective}. As Weizenbaum warned decades ago \cite{weizenbaum1976computer}, AI lacks the capacity for genuine relationship---it cannot care about students' growth, model professional identity, or provide the emotional scaffolding that sustains learning through difficulty. A substantial body of research shows that learning is fundamentally social. Vygotskian perspectives emphasize that reasoning develops through interaction, dialogue, and scaffolded participation in authentic practice, where students learn to exercise situated judgment rather than merely follow procedures \cite{vygotsky1978mind,johri2011situated}.

\subsection{The Problem with Individualistic AI}
Despite robust evidence for social learning's effectiveness, GenAI implementation in engineering classrooms often is individualistic (e.g., students consulting ChatGPT or custom GPTs privately, receiving personalized explanations, generating solutions in isolation \cite{qadir2025generative}). This has the following drawbacks.

\subsubsection{Elimination of Peer Interaction}
When students turn to AI rather than classmates, they lose opportunities for explaining concepts to others (which deepens understanding), encountering alternative solution approaches, negotiating meaning through constructive disagreement, and developing communication skills essential for engineering practice.

\subsubsection{Removal of Productive Struggle}
AI bypasses the ``desirable difficulty'' \cite{bjork2011making} that comes from grappling with information and alternatives, a struggle essential for deep understanding. This creates what Sarofian-Butin dubs ``\textit{kayfabe}", borrowing professional wrestling's term for staged events presented as real, where students produce academic outputs without the cognitive work that generates genuine understanding \cite{sarofianbutin2025kayfabe}. 

\subsubsection{Creation of Dependency}
In the MIT Media Lab study \cite{kosmyna2025cognitive}, students using ChatGPT for essay writing displayed significantly weaker brain connectivity than those working independently. When later required to write without AI, these students exhibited persistently reduced neural engagement compared to peers who had never used ChatGPT, struggled to quote sentences from their own essays, and reported the lowest sense of ownership over their work.

\subsubsection{Lack of Authentic Context}
Engineering problems in practice emerge from messy real-world contexts, involve multiple stakeholders with competing priorities, require navigating ambiguity and incomplete information. Individual AI tutoring abstracts away this complexity, presenting sanitized technical puzzles divorced from the socio-technical systems thinking engineering demands.

\subsubsection{Obscuring of Thinking Processes}
AI interaction privatizes learning, eliminating the pedagogical visibility that makes collaborative environments educationally rich. When students work together, they externalize reasoning and expose misconceptions---thinking becomes observable to teachers and peers, promoting the development of CI. 

\subsection{Education as Divergent Ongoing Problem}
Not all problems can be solved---some must be continually managed. Schumacher distinguished between \textit{convergent problems}, which yield to definitive solutions through technical means, and \textit{divergent problems}, which involve balancing competing goods that remain perpetually in tension \cite{schumacher1977guide}. Education represents a quintessential divergent problem: it requires simultaneously honoring multiple valid yet contradictory imperatives, none of which can be permanently resolved in favor of the others. Polarity management theory frames this as managing ongoing tensions rather than solving discrete challenges \cite{qadir2024learning}. The introduction of GenAI intensifies the following three core educational polarities:

\begin{itemize}

\item \textit{Individual Mastery vs. Collaborative Capability}: Assessment measures individual achievement to ensure each student develops foundational competencies, yet professional engineering operates through distributed cognition where teams integrate diverse expertise and build solutions no individual could create alone. AI tutoring optimizes personal learning but eliminates practice in articulating reasoning to peers and negotiating design trade-offs---both individual mastery and collaborative capability remain irreducibly essential.

\item \textit{Knowledge Transmission vs. Knowledge Construction}: AI excels at efficient information delivery through clear explanations and immediate answers, yet constructivist theory demonstrates that understanding cannot be deposited but must be actively constructed through struggle and personal sense-making \cite{vygotsky1978mind}. Freire warned that treating students as information receptacles dehumanizes learning \cite{freire1970pedagogy}; AI risks amplifying this ``banking education" at scale while students need both efficiently transmitted knowledge and cognitively demanding space to construct transferable understanding.

\item \textit{Cognitive Development vs. Social-Emotional Growth}: Engineering curricula prioritize technical mastery as the foundation of professional competence, yet practice simultaneously demands social-emotional capacities: communicating with stakeholders, navigating ethical dilemmas, and exercising empathy in design. AI tutoring optimizes cognitive skill acquisition but offers no practice in the relational and ethical dimensions of responsible engineering. However, both technical depth and social-emotional capability remain essential for meaningful professional contribution.
\end{itemize}

These polarities cannot be resolved by choosing one pole over another as both poles are essential. Individualistic AI interaction optimizes efficiency and immediate performance while sacrificing deep learning, collaboration, and social-emotional development. Collective intelligence approaches attempt to hold both poles in creative tension, leveraging AI's capabilities while preserving human reasoning, social interaction, and the developmental struggles that build expertise.

\subsection{Timing and Orchestration in Human-AI Collaboration}

Recent research demonstrates that \textit{how} and \textit{when} AI is integrated matters as much as \textit{whether} AI is used. Agarwal et al. \cite{agarwal2023combining} found that the timing of AI input relative to human decision-making significantly impacts outcomes in the field of healthcare (radiology), while Vaccaro et al.'s \cite{vaccaro2024combinations} meta-analysis identifies conditions under which human-AI combinations prove useful versus where they are counterproductive. Burton et al. \cite{burton2024large} emphasize that LLM effectiveness depends on deliberate design choices rather than inherent technological properties---a principle aligning with our pedagogical approach requiring careful orchestration of when groups consult AI, how prompts are structured, and how outputs are critically evaluated. This literature suggests that the common pattern of students consulting AI immediately when encountering difficulty represents potentially the least effective timing. Alternative approaches include productive failure followed by AI consultation, initial human collaboration, then AI critique, or AI-generated options as starting points for human evaluation. Our implementation experiments with these timing patterns, seeking empirical evidence for optimal orchestration in engineering education contexts.

\subsection{Cultural Bias in LLMs and the Need for Diversity}

A critical concern for engineering education's global context involves cultural biases embedded in large language models (LLMs). Our recent research \cite{mushtaq2025towards,mushtaq2025worldview} demonstrates that frontier LLMs exhibit systematic biases reflecting primarily Western, Anglo-American worldviews. These AI models often present perspectives that may conflict with local cultural values, oversimplify or stereotype cultures, or present Western approaches as universal rather than culturally situated. This represents a form of epistemic injustice---students' own cultural knowledge and ways of knowing become marginalized or invisible. CI approaches provide partial mitigation: when students reason together first, they activate local knowledge, cultural perspectives, and contextual understanding before AI consultation. Group discussion allows students to evaluate AI outputs against their lived experience, identify biases or cultural misalignments, and interrogate assumptions embedded in AI-generated content. This positions students as active knowledge creators who critically assess and selectively appropriate AI suggestions rather than passive recipients of potentially biased information \cite{mushtaq2025towards}.

\section{Generative Collective Intelligence}
\label{sec:GCI}

\subsection{Making Thinking Visible: Thinking Routines}

Harvard Project Zero's Making Thinking Visible framework \cite{ritchhart2020power} provides proven structures for making student thinking visible, fostering collaborative sense-making, and developing intellectual dispositions. Grounded in principles that learning is a consequence of thinking, that thinking is dispositional (requiring curiosity, open-mindedness, attention to evidence), and that development of thinking is fundamentally social, these routines transform classrooms from work-oriented to learning-oriented cultures. Thinking routines are small sets of questions or brief sequences of steps that scaffold and support thinking. Through repeated use, they become ``patterns of thinking" woven into classroom culture, making cognition visible to both teachers (for assessment and understanding) and learners (for metacognitive awareness). The key distinction from algorithmic problem-solving lies in routines' open-ended, dialogical nature---they structure inquiry without prescribing answers, supporting independence rather than creating dependence.

\vspace{0.1mm}
\subsubsection{Question Sorts}

Question Sorts thinking routine \cite{ritchhart2020power} exemplifies thinking routines' power for collaborative inquiry. Students first generate questions in response to a shared stimulus. They then sort these questions along two axes: \textit{generativity}---the extent to which a question sparks engagement, insight, or creative exploration---and \textit{genuineness}---how much the question truly matters to them and is worth investigating further. In engineering contexts, Question Sorts supports problem framing in design projects, identifies knowledge gaps in complex systems, surfaces diverse perspectives on technical challenges, and prioritizes group investigation directions. Unlike AI-generated question lists that students passively consume, Question Sorts requires active evaluation, collaborative negotiation, and shared ownership of inquiry direction---making it fundamentally a collective intelligence activity rather than individual consumption.

\vspace{0.1mm}
\subsubsection{Peel the Fruit}

Peel the Fruit thinking routine \cite{ritchhart2020power} provides a complementary structure for exploring topics from surface observations to core principles. This planning and tracking map guides learners through layers: \textit{the skin} (surface observations and initial familiarity), \textit{the substance} (parts and perspectives, connections, reasons and explanations, changes over time), and finally \textit{the core} (central insights and essential truths). For example, for Internet ethics specifically, students observe \textit{surface technical facts} (IP addresses identify devices), investigate \textit{underlying mechanisms} (central registries assign blocks based on what principles?), explore \textit{substantive connections} (who controls allocation and why?), and ultimately reach \textit{core ethical implications} (how technical architecture may create digital inequality and embeds power relations).

\subsection{The GCI Framework}

Our pedagogical approach draws directly from Kehler, Page, Pentland, Reeves, and Brown's 2025 Generative Collective Intelligence (GCI) framework \cite{kehler2025amplifying}, which reconceptualizes AI as ``a new kind of cultural and social technology" creating a \textit{cognitive bridge} between human reasoning and computational capabilities. This framework rejects the view of intelligence as an isolated, algorithmic property and instead understands it as fundamentally social, relational, and collectively constructed. GCI positions AI in dual roles: as an interactive agent providing computational power AND as a technology that accumulates and organizes collective knowledge through structured collaboration. Our classroom implementation operationalizes GCI principles through carefully sequenced thinking routine activities. Students begin with Question Sorts, generating diverse questions about Internet ethics through collaborative inquiry without AI assistance---mirroring GCI's emphasis on human reasoning establishing the foundation. Next, Peel the Fruit structures collective analysis as groups trace connections from technical architecture through social consequences to ethical implications, building what GCI calls ``conceptual blending" across mental spaces. Only after this human-centered reasoning do we introduce strategic AI consultation at two moments: first, to surface additional perspectives; second, to critique and extend preliminary analyses. This sequence embodies GCI's core principle: AI enhances rather than replaces collective human intelligence. Students experience AI not as oracle but as one voice among many, to be interrogated and integrated through collective judgment.

\subsection{Interactionalism for Higher Education}

Moldoveanu and Siemens' 2025 interactionalism framework \cite{moldoveanu2025interactionalism} highlights a core shift: \textit{skills are now exercised interactionally and dialogically before they are individually mastered}. This shift requires cultivating \textit{interactional intelligence}, which integrates \textit{meta-cognitive capacities} (understanding how we think \textit{with} AI, evaluating model appropriateness, and designing task architectures that distribute work between human and machine) and \textit{meta-emotional capacities} (managing frustration and over-confidence, calibrating trust, and maintaining engagement when AI handles routine labor). Because over 80\% of modern work occurs in groups and social–relational skills have outpaced cognitive skills in importance for more than two decades, the framework argues that AI-enabled learning must move beyond isolated tutoring toward \textit{continuous, dialogical, personalized support embedded within collective learning contexts}. In this view, AI augments---rather than replaces---the fundamentally social processes through which students learn to code, troubleshoot networks, design systems, and build professional judgment.

\section{Methodology}
\label{sec:methodology}

Our dual-pronged approach integrates theoretical synthesis with empirical classroom implementation, as detailed next.

\subsection{Literature Review and Theoretical Synthesis}

We synthesize insights from multiple related fields including learning sciences (social constructivism, active learning, cooperative learning, scaffolding theory, productive failure \cite{kapur2008productive}), collective intelligence theory (GenCI framework \cite{kehler2025amplifying}, interactionalism \cite{moldoveanu2025interactionalism}), embodied cognition and phenomenology \cite{dreyfus1992computers,varela1991embodied,winograd1986understanding}, philosophy of technology (values in design, ethics of AI, critical perspectives on educational technology \cite{vallor2022oxford}), human-AI interaction research including design guidelines \cite{amershi2019guidelines,sellen2024rise} and studies on timing and orchestration \cite{agarwal2023combining,vaccaro2024combinations,burton2024large}, cultural bias in generative AI \cite{mushtaq2025towards,mushtaq2025worldview}, and engineering education scholarship including previous work on GenAI integration \cite{qadir2023engineering,qadir2025generative,johri2023generative, qadir2024learning}. This synthesis generated insights about requirements for effective GCI pedagogy: maintaining social learning dimensions, strategic timing of AI consultation, making thinking visible, developing metacognitive awareness, addressing cultural biases, and preparing for interactional rather than purely individual skill exercise.

\subsection{Our GCI Pedagogical Framework}

We synthesize GCI, interactionalism, and thinking routines into a coherent pedagogical approach for engineering education. This represents one possible implementation among many, designed specifically for our context and constraints.

\subsubsection{Core Design Principles}

Our framework is based on the following five GCI elements in classroom practice: 

\begin{itemize}
 
\item \textit{Human reasoning leads}---activities begin with individual thinking then pair/group collaboration using thinking routines to structure inquiry, generate questions, and analyze concepts from multiple layers, establishing collective foundation before AI consultation; 
\item \textit{Strategic AI consultation}---only after human reasoning establishes foundation do groups consult AI for specific purposes (surfacing additional perspectives, critiquing preliminary analyses, generating alternative approaches); 
\item \textit{Critical evaluation}---groups interrogate AI outputs against collective human reasoning, explicitly identifying what AI adds and misses, synthesizing contributions rather than uncritically accepting outputs; 
\item \textit{Artifact creation}---groups produce tangible records (concept maps, diagrams, synthesis frameworks) capturing evolving collective knowledge construction and providing visible evidence of thinking processes; and 
\item \textit{Metacognitive reflection}---activities conclude with explicit reflection examining what humans contribute that AI cannot, when AI adds genuine value, and how collective intelligence emerges.

\end{itemize}

\subsubsection{Rationale for Design Choices}

This approach recognizes AI's complementary rather than competitive relationship with human intelligence. AI cannot engage in authentic ethical reasoning, lacks embodied understanding of real-world contexts, exhibits cultural biases, or form genuine relationships---but can rapidly synthesize information, generate alternative framings, and facilitate a comparison of multiple approaches. Our implementation addresses individualistic AI's limitations by maintaining productive struggle and peer interaction (countering cognitive debt), making thinking visible (enabling formative assessment), developing interactional intelligence (preparing for AI-augmented workplaces), and creating equitable learning environments through collective learning based on diverse perspectives.

\subsection{Classroom Implementation}

We implemented GCI activities across two undergraduate engineering courses at Qatar University during Fall 2025 semester. The first course, Data Communications and Computer Networks I (CMPE355), or \textbf{DCCN-I} , taught by the first author in the Computer Engineering program, focused on Ethics of Internet and Digital Inequality---examining how Internet design embeds ethical choices and tracing connections from technical architecture to social consequences. This course had 94 female students in two sections (size of 50 and 44, respectively)---with 68 students filling the questionnaire that contributed to this study. The second course, Signals and Systems (ELEC351), or \textbf{S\&S}, taught by the second author in the Electrical Engineering program, implemented parallel activities with domain-specific content. This course had 89 mixed students in two sections (size of 36 (male) and 53 (female), respectively)---with 72 students filling the questionnaire. Both courses used an identical pedagogical structure inspired by GCI principles and the thinking routines framework across two 50-minute class sessions (100 minutes total).

\subsubsection{Activity Sequence and Design}

Activities progressed through seven phases designed to operationalize our GenCI framework. First, individual warm-up prompts asked provocative questions to activate prior knowledge. Following this, pairs engaged in \underline{Question Sorts} by generating questions about the topic, then groups of 4-5 sorted them into two principal categories---\textit{clarifying} (factual) and \textit{generative} (higher-order thinking exploring implications, consequences or possibilities)---identifying their top two thought-provoking questions in each category. In the first strategic AI consultation phase, each group used GenAI (ChatGPT, and for DCCN-I, a CustomGPT) to explore one of their top questions with the required prompt structure ``\textit{Summarize multiple perspectives on}..." to surface diverse viewpoints rather than definitive answers. Subsequently, groups engaged in \underline{Peel the Fruit} by choosing one focal concept and collaboratively creating layered diagrams showing \textit{outer layer} (surface technical facts), \textit{middle layer} (underlying mechanisms and control), and \textit{core} (human/ethical consequences regarding power, access, fairness). The second strategic AI consultation asked groups to critique or extend their fruit diagram (``\textit{What ethical issues or real-world examples are missing from our analysis}?"), noting perspectives that added value or challenged thinking. Following group work, instructor-facilitated synthesis drew connections across group insights. Finally, individual exit tickets prompted reflection: ``\textit{How did our group's collective intelligence change after consulting GenAI?}"

\subsubsection{Implementation Principles}

Several design principles guided implementation. We explicitly defined GCI for students as enhanced capacity emerging when we think together (collective intelligence), creating new insights not just aggregating existing ones (generative), through strategically consulting AI to enhance rather than replace human reasoning (human + AI). We contrasted our model with traditional individual AI use: instead of individuals searching AI, getting answers, and moving on (producing shallow engagement where ethics stay invisible), our approach prioritized collective human intelligence first, then strategic AI consultation, enabling deeper synthesis. Timing proved critical: two carefully-timed AI consultations occurred after human question generation but before deep analysis, then after human analysis but before final synthesis---ensuring human reasoning established foundation and critically evaluated AI contributions rather than AI framing initial thinking. Throughout activities, we emphasized metacognitive framing: students were developing awareness of their own thinking processes, learning to work productively with AI rather than depend on it, and building capacity for critical evaluation of AI outputs.

\subsection{Data Collection}

We collected the data in the following ways:

\subsubsection{Comprehensive Questionnaire}

A comprehensive questionnaire was administered to the students during the second activity session, through which both quantitative and qualitative data were gathered. The survey included 16 closed-ended items---10 Likert-scale questions assessing the effectiveness of different learning approaches, the perceived value of AI versus human collaboration, confidence levels, and quality of learning outcomes, along with 6 multiple-choice questions examining preferred learning methods, frequency of GenAI use, optimal timing for AI consultation, and concerns about over-reliance. In addition, 7 open-ended questions probed students' detailed experiences with different learning modalities, examples of effective AI use, concerns about limitations and biases, suggestions for improvement, reflections on collective versus individual learning, and ethical considerations. 

\subsubsection{Instructor Observations}

Both instructors maintained observational notes documenting student engagement levels during different activity phases, quality of questions generated during Question Sorts, depth of analysis in Peel the Fruit diagrams, nature of AI consultations, including prompt quality and critical evaluation of outputs, group dynamics and collaboration patterns, and comments during reflection phases.



\subsection{Analysis Approach}

Survey data were analyzed using descriptive statistics to examine patterns in effectiveness ratings, AI usage frequencies, and timing preferences. Open-ended responses underwent thematic analysis with Microsoft Co-Pilot (institutionally supported), assisting in identifying initial patterns and suggesting potential themes from coded data. We then refined these themes and connected them to our research questions and theoretical frameworks. Major themes identified included the value of human collaboration, risks of AI over-reliance, importance of timing in AI consultation, AI as a perspective-broadener, metacognitive awareness development, and cultural and ethical concerns. Quantitative patterns and qualitative themes were integrated through mixed-methods synthesis, with instructor observations providing additional triangulation.

\section{Results}
\label{sec:results}

\subsection{Learning Method Effectiveness}

\begin{figure}[h]
\centering
\begin{tikzpicture}
\begin{axis}[
 ybar,
 width=0.85\columnwidth,
 height=5.5cm,
 ylabel={Percentage of Students (\%)},
 symbolic x coords={Studying alone, Using AI alone, Group work (no AI), Group work + AI},
 xtick=data,
 x tick label style={rotate=35,anchor=east,font=\small},
 ymin=0,
 ymax=70,
 bar width=10pt,
 nodes near coords,
 nodes near coords align={vertical},
 every node near coord/.append style={font=\scriptsize},
 nodes near coords style={/pgf/number format/.cd,fixed,precision=1,/tikz/.cd},
 legend style={at={(0.02,0.98)},anchor=north west,legend columns=1,font=\small},
 legend image code/.code={%
 \draw[#1] (0cm,-0.1cm) rectangle (0.3cm,0.1cm);
 },
]
\addplot[fill=blue!70!black] coordinates {
 (Studying alone,19.1)
 (Using AI alone,44.1)
 (Group work (no AI),2.9)
 (Group work + AI,33.8)
};
\addplot[fill=teal!70!black] coordinates {
 (Studying alone,13.9)
 (Using AI alone,15.3)
 (Group work (no AI),5.6)
 (Group work + AI,65.3)
};
\legend{DCCN-I ($N=68$), S\&S ($N=72$)}
\end{axis}
\end{tikzpicture}
\caption{Student preferences for learning methods. ``Group work + AI (GCI)" was most popular (50\% overall). DCCN-I students with prior custom chatbot access showed higher individual AI preference (44\%) versus S\&S (15\%).}
\label{fig:learning_methods}
\end{figure}
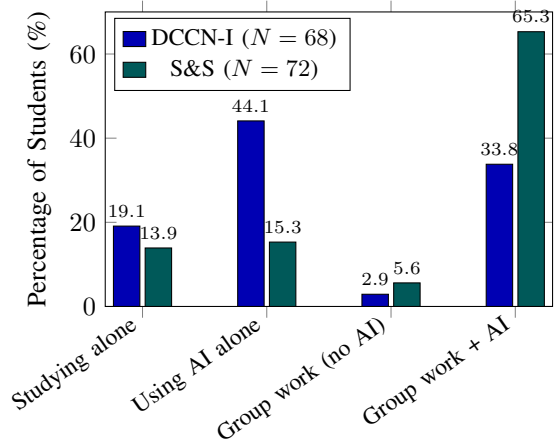

Students' preferences for different learning modalities revealed clear patterns regarding individual versus collective approaches and the role of AI support. Figure \ref{fig:learning_methods} shows that ``Group work + AI (GCI)" was most popular overall (50\%), followed by ``Using AI alone" (29\%), ``Studying alone" (16\%), with ``Group work without AI" least preferred (4\%). However, striking differences emerged between courses: DCCN-I students showed higher preference for individual AI use (44\%) compared to S\&S students (15\%), while S\&S students strongly favored GCI (65\%) versus DCCN-I (34\%).

These results reveal important nuances requiring careful interpretation. DCCN-I students' higher preference for individual AI use likely reflects their extensive prior experience with a custom chatbot available throughout the course, while both cohorts experienced structured GCI activities for only two 50-minute sessions. Preferences naturally reflect familiarity. That S\&S students strongly favored GCI (65\%) after minimal exposure suggests significant potential when students lack established individual AI habits. The near-absence of preference for group work without AI (only 4\% overall) may reflect implementation design---we contrasted individual AI use with strategic collective AI consultation rather than providing extensive group work without any AI access for comparison. 

Qualitative data (discussed below) shows students recognizing limitations of individual AI use even while rating it effective---awareness of risks around over-reliance, bias, and shallow engagement that quantitative preferences alone do not capture.

\subsection{GenAI Usage Frequency}

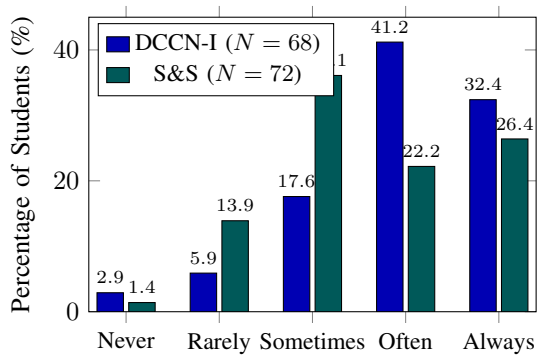
\begin{figure}[htbp]
\centering
\begin{tikzpicture}
\begin{axis}[
 ybar,
 width=0.85\columnwidth,
 height=5.5cm,
 ylabel={Percentage of Students (\%)},
 symbolic x coords={Never, Rarely, Sometimes, Often, Always},
 xtick=data,
 x tick label style={font=\small},
 ymin=0,
 ymax=45,
 bar width=10pt,
 nodes near coords,
 nodes near coords align={vertical},
 every node near coord/.append style={font=\scriptsize},
 nodes near coords style={/pgf/number format/.cd,fixed,precision=1,/tikz/.cd},
 legend style={at={(0.02,0.98)},anchor=north west,legend columns=1,font=\small},
 legend image code/.code={%
 \draw[#1] (0cm,-0.1cm) rectangle (0.3cm,0.1cm);
 },
]
\addplot[fill=blue!70!black] coordinates {
 (Never,2.9)
 (Rarely,5.9)
 (Sometimes,17.6)
 (Often,41.2)
 (Always,32.4)
};
\addplot[fill=teal!70!black] coordinates {
 (Never,1.4)
 (Rarely,13.9)
 (Sometimes,36.1)
 (Often,22.2)
 (Always,26.4)
};
\legend{DCCN-I ($N=68$), S\&S ($N=72$)}
\end{axis}
\end{tikzpicture}
\caption{Frequency of GenAI tool usage across DCCN-I (blue) and S\&S (teal) courses. Overall, 61\% of students used AI ``Often" or ``Always", indicating substantial baseline engagement with AI tools.}
\label{fig:ai_frequency}
\end{figure}

Students reported their frequency of using generative AI tools (CustomGPT, ChatGPT, etc.) during the course as shown in Figure \ref{fig:ai_frequency}. 

This figure reveals extensive GenAI adoption: 61\% of students reported using AI tools ``Often" or ``Always", with notable variation between courses. DCCN-I students, who had access to a custom chatbot throughout the course, showed higher frequent usage (74\% used AI ``Often" or ``Always") compared to S\&S students (49\%). Overall, only 12\% used AI ``Rarely" or ``Never". This high baseline usage validates our study's relevance---these are not students unfamiliar with AI who might be overwhelmed by its introduction, but rather regular AI users for whom our question concerns optimal integration approaches. The prevalence of ``Often" and ``Always" usage also contextualizes concerns about over-reliance. With three-quarters of students using AI frequently, pedagogical approaches that ignore AI or attempt to prevent its use appear unrealistic. The question becomes not whether students use AI but how to structure that use productively.

\subsection{Timing Effects on Learning}

\begin{figure}[h]
\centering
\begin{tikzpicture}
\begin{axis}[
 ybar,
 width=0.85\columnwidth,
 height=5.5cm,
 ylabel={Percentage of Students (\%)},
 symbolic x coords={Big difference, Sometimes matters, Doesn't matter, Not sure},
 xtick=data,
 x tick label style={rotate=25,anchor=east,font=\small},
 ymin=0,
 ymax=55,
 bar width=10pt,
 nodes near coords,
 nodes near coords align={vertical},
 every node near coord/.append style={font=\scriptsize},
 nodes near coords style={/pgf/number format/.cd,fixed,precision=1,/tikz/.cd},
 legend style={at={(0.98,0.98)},anchor=north east,legend columns=1,font=\small},
 legend image code/.code={%
 \draw[#1] (0cm,-0.1cm) rectangle (0.3cm,0.1cm);
 },
]
\addplot[fill=blue!70!black] coordinates {
 (Big difference,51.5)
 (Sometimes matters,39.7)
 (Doesn't matter,5.9)
 (Not sure,2.9)
};
\addplot[fill=teal!70!black] coordinates {
 (Big difference,47.2)
 (Sometimes matters,43.1)
 (Doesn't matter,4.2)
 (Not sure,5.6)
};
\legend{DCCN-I ($N=68$), S\&S ($N=72$)}
\end{axis}
\end{tikzpicture}
\caption{Impact of AI consultation timing on learning. Overall, 90\% of students believe timing makes at least some difference, with approximately half stating it makes a ``big difference".}
\label{fig:timing_impact}
\end{figure}
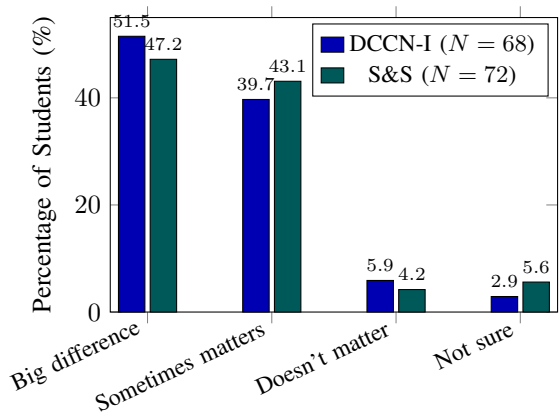

A critical finding concerned whether timing of AI consultation relative to human reasoning affects learning effectiveness. Figure \ref{fig:timing_impact} shows strong student recognition that timing matters across both courses: approximately half of students (49\%) indicated timing of AI consultation (before, during, or after group work) makes a ``big difference" to learning effectiveness, while an additional 41\% said it matters ``sometimes". Only 5\% believed timing doesn't matter, with 4\% unsure. Overall, 90\% of students recognize that timing makes at least some difference, demonstrating sophisticated awareness that \textit{when} they engage AI significantly affects learning outcomes.

This finding proves particularly significant given recent research on human-AI collaboration demonstrating timing and orchestration effects \cite{agarwal2023combining,vaccaro2024combinations}. Students' intuitive recognition that when they consult AI matters aligns with evidence from other domains showing that immediate AI consultation can prevent productive struggle while delayed consultation can extend rather than replace human reasoning.

Our structured approach---human reasoning first through Question Sorts and initial Peel the Fruit analysis, then strategic AI consultation at two defined points---appears to have made timing's importance salient to students. This metacognitive awareness represents a learning outcome itself: students developing judgment about when AI consultation adds versus detracts from learning.

\subsection{Thematic Analysis of Open-Ended Reflections}

Thematic analysis of open-ended responses revealed four recurring themes illuminating students' nuanced understanding of AI integration in learning, with notable differences between courses reflecting their distinct experiences with AI tools. Student quotes have been lightly edited for grammar while preserving original meaning.

\subsubsection{\textbf{Theme 1 (Synergy between human intelligence \& AI)}}

Students recognized complementary strengths of human reasoning and AI capabilities. S\&S students emphasized unique human contributions:

\begin{quote}
\textit{``AI made us faster, but our discussions made us smarter.''}
\end{quote}

\begin{quote}
\textit{``AI sometimes missed emotional cues or subtle meanings that were obvious to us.''}
\end{quote}

\begin{quote}
\textit{``collective human intelligence builds up communication skills and understanding each other's ideas that are based in each one's life experience, rather than rigid data and information''}
\end{quote}

DCCN-I students noted how GCI deepened understanding beyond individual AI use:

\begin{quote}
\textit{``GCI activities really helped me understand DCCN-I beyond just the lectures... It also pushed me to think critically, solve problems on my own, and collaborate with others.''}
\end{quote}

\begin{quote}
\textit{``I actually learned a lot from this session... working in groups helped me understand how others think.''}
\end{quote}

\subsubsection{\textbf{Theme 2 (Critical awareness of over-reliance risks)}}

DCCN-I students articulated over-reliance from personal experience, often in confessional tones:

\begin{quote}
\textit{``I relied too much on AI for the assignments before the midterm... after the midterm, I decided to use it more strategically as a tool rather than giving it most of the work to do.''}
\end{quote}

\begin{quote}
\textit{``over-relying on AI instead of using my brain and developing my own critical thinking''}
\end{quote}

S\&S students framed over-reliance as a future risk requiring active prevention:

\begin{quote}
\textit{``So people don't use to just rely on it, this will make our minds weaker cause it's like any muscle''}
\end{quote}

\begin{quote}
\textit{``Not letting the human brain shut down completely.''}
\end{quote}

\begin{quote}
\textit{``To not depend on AI fully but to try to use our intelligence instead.''}
\end{quote}

This pattern suggests GCI pedagogy serves dual purposes: helping students already experiencing over-reliance develop strategic awareness while preventing dependency from developing in less experienced users.

\subsubsection{\textbf{Theme 3 (Bias, accuracy, and critical evaluation)}}

DCCN-I students provided specific examples of bias:

\begin{quote}
\textit{``The examples chatGPT gives are clearly biased. For example... it only provided Iran and Turkmenistan, not Israel which has frequent telecom, internet shutdowns in Gaza.''}
\end{quote}

\begin{quote}
\textit{``Sometimes AI gave incorrect or unclear explanations, which could confuse the group... we had to fact-check and balance its use.''}
\end{quote}

S\&S students emphasized group verification as strength:

\begin{quote}
\textit{``correction is the best form of learning, when using AI you don't have access to this method of learning whereas in a group you get access to different sorts of thinking and feedback''}
\end{quote}

\begin{quote}
\textit{``AI gives a straight answer, in groups we could discuss more.''}
\end{quote}

Both cohorts demonstrated emerging critical AI literacy---recognizing systemic limitations while strategically leveraging capabilities.

\subsubsection{\textbf{Theme 4 (Strategic AI use and human-first reasoning)}}

DCCN-I students articulated learned wisdom about timing:

\begin{quote}
\textit{``I try to not entirely rely on AI and think before going to AI because it's always easier to go to AI than do the thinking but that's a disadvantage.''}
\end{quote}

\begin{quote}
\textit{``The activity taught me that AI is a tool for exploring ideas, not an authority that gives truth.''}
\end{quote}

S\&S students recognized this principle through structured activities:

\begin{quote}
\textit{``We learned to use AI as support while our group did the real reasoning.''}
\end{quote}

\begin{quote}
\textit{``AI has more information, people have experience''}
\end{quote}

This reflective maturity exemplifies GCI pedagogy's goal: students discerning appropriate boundaries for AI use---with S\&S students showing immediate appreciation for collective approaches while DCCN-I students demonstrated wisdom from experiencing over-reliance consequences firsthand.

\subsection{Instructor Observations}

Both instructors noted high engagement during Question Sorts and Peel the Fruit activities, with energy levels noticeably exceeding typical lecture-based sessions. Question Sorts generated remarkably diverse and sophisticated questions beyond mere factual clarification, with groups producing higher-order thinking exploring ethical implications and imagining alternatives---diversity challenging to achieve individually or with AI alone. Peel the Fruit diagrams showed students progressing from surface technical facts to nuanced understanding of power, control, and ethical consequences, with many groups making sophisticated connections between technical architecture and social implications that likely would not emerge from individual AI consultation. During AI consultation phases, groups engaged critically rather than passively accepting outputs, comparing AI responses with their pre-existing analysis, debating whether AI-suggested perspectives were valid, identifying specific additions AI made versus merely restating known information, and noticing biases or limitations in outputs. 

\subsection{Limitations}

Significant limitations prevent strong causal claims and qualify conclusions. Internal validity is constrained by absence of control groups, no random assignment, potential novelty effects, instructor enthusiasm potentially biasing observations, and self-report data that may reflect social desirability rather than genuine perceptions. External validity is limited by single institution implementation in Qatar, small sample size (140 students across two related courses), voluntary participation potentially selecting motivated students, and short duration (two 50-minute sessions). Construct validity concerns include non-validated survey items developed for this study, subjective perceptions that may not correlate with objective learning outcomes, and multiple possible operationalizations of ``effectiveness" beyond what we measured. Reliability is threatened by first-time implementation with inevitable variations, different instructors across courses despite same design, and qualitative coding by single researcher without inter-rater reliability. Students' preferences for individual AI use may reflect familiarity and immediate efficiency rather than superior learning outcomes. This exploratory work establishes proof-of-concept requiring validation through randomized controlled trials, objective learning outcomes, longitudinal studies tracking retention and transfer, cross-cultural replication, and standardized implementation protocols.

\section{Discussion and Future Directions}
\label{sec:discussion}

\subsection{Interpreting Student Preferences}

The apparent contradiction between quantitative preferences (44\% favoring individual AI use) and qualitative recognition of collective learning's advantages reflects that students facing heavy workloads naturally gravitate toward approaches minimizing immediate cognitive load, even while recognizing these may not maximize learning depth. Individual AI consultation provides instant gratification---quick answers, reduced struggle, completion satisfaction---while collective learning's benefits manifest later and remain less salient in immediate preference reporting. That one-third of students found GCI comparably effective after only two sessions suggests significant potential as familiarity increases. Critically, what students prefer is not necessarily what enables deepest learning, just as students might prefer lectures over active learning yet learn more from active approaches. Future research requires objective performance measures comparing learning gains across modalities, examining whether GCI produces the adaptive, transfer-ready knowledge that individual AI tutoring cannot.

\subsection{The Critical Role of Timing}

Students' recognition that timing matters represents a key finding with both theoretical and practical implications, aligning with recent human-AI collaboration research showing that when AI input occurs relative to human decision-making significantly affects outcomes. Immediate AI consultation can anchor thinking to AI framings and prevent productive struggle, while delayed consultation allows humans to develop initial framings and engage in productive failure before AI extends rather than replaces reasoning. Our approach---human-first reasoning through thinking routines, followed by two strategic AI consultations at defined points---made timing salient and cultivated metacognitive awareness about when AI adds versus detracts value. Future research should systematically compare timing patterns (\textit{AI-first}, \textit{AI-concurrent}, \textit{AI-delayed}, \textit{AI-iterative}, and \textit{AI-final}) to identify optimal orchestration patterns for different learning objectives and student populations, recognizing that strategic timing may be as important as the AI tools themselves.

\subsection{Toward Multi-Agent Collective Intelligence Systems}

The most promising direction extends beyond isolated AI consultations toward multi-agent systems that facilitate genuine collective intelligence at scale. Rather than each student consulting a separate AI instance, future implementations could employ agentic AI systems using protocols like Model Context Protocol (MCP) to orchestrate collective sense-making: aggregating student questions into maps revealing knowledge gaps, simulating diverse stakeholder perspectives, facilitating comparative judgment across groups, documenting idea provenance, and identifying emergent patterns no individual could detect.

\subsubsection{Conversational Swarm Intelligence}

Recent advances in conversational swarm intelligence demonstrate how real-time deliberation among humans and AI agents can amplify collective reasoning beyond what either achieves alone \cite{rosenberg2025large}. Unlike traditional brainstorming that aggregates individual opinions, conversational swarms enable synchronous collective reasoning where participants influence each other through rapid exchanges, creating emergent insights through dialectical synthesis \cite{rosenberg2025collective}. Applied to engineering education, swarms could enable distributed student groups to collectively design network architectures, analyze signal processing trade-offs, and evaluate ethical implications---with AI agents facilitating dynamics, surfacing overlooked perspectives, and helping groups converge on robust solutions.

\subsubsection{Implementation Pathways}

Platforms like \url{Deliberation.io} (developed at Stanford HAI and MIT) demonstrate how structured protocols employing anonymized contribution, adaptive idea selection, and transparent provenance tracking can harness collective intelligence while maintaining individual agency. Educational implementations could support both synchronous swarms for real-time collaborative reasoning as well as asynchronous deliberation where students across institutions collaborate on extended engineering challenges. This represents a fundamental shift from AI as individual tutor to AI as collective intelligence orchestrator---amplifying human creativity through structured collaboration while preserving human judgment and responsibility.

\subsection{Implementation Guidance and Research Priorities}

For educators implementing GCI approaches, essential principles include starting with single activities, making pedagogical rationale explicit, providing substantial scaffolding, ensuring psychological safety, timing activities realistically, documenting collective artifacts, and most critically, aligning assessment with pedagogy---if evaluation remains purely individual, students rationally prioritize individual preparation over collective learning. Future research must address current limitations through randomized controlled trials with objective outcomes, longitudinal studies tracking effects into professional practice, mechanism studies isolating which elements drive learning gains, cross-cultural replications establishing generalizability, and multi-agent system research exploring how AI orchestrators can facilitate collective reasoning across distributed learning communities---representing a promising frontier that could transform how engineering knowledge is collectively constructed across institutions.

\section{Conclusions}
\label{sec:conclusion}

Generative AI's integration into engineering education risks reinforcing individualistic learning that optimizes efficiency at the expense of deep understanding. This paper argues for an alternative approach: harnessing AI to catalyze collective intelligence that amplifies human creativity and collaborative capacity. We develop a Generative Collective Intelligence (GCI) framework integrating Harvard Project Zero's Making Thinking Visible routines with strategic AI consultation, positioning AI as a cognitive bridge rather than a replacement tutor. Our implementation through structured activities---Question Sorts and Peel the Fruit combined with strategic AI consultation---yields four key insights: students demonstrate sophisticated awareness of both AI's value and risks, moving beyond simplistic positions; \textit{timing matters critically}, with students recognizing intuitively that when they engage AI significantly affects learning outcomes; human collaboration provides irreplaceable benefits---deeper understanding, creative questioning, critical challenge---that individual AI use cannot replicate; and structured thinking routines cultivate metacognitive awareness and appropriate AI calibration. However, the lack of controls, small sample, single context, short duration, and self-report measures limit interpretability, making this an exploratory proof-of-concept requiring rigorous follow-up. The question is not \textit{whether} AI enters classrooms but \textit{how}---whether it reinforces isolation or fosters collective mastery, produces cognitive atrophy or amplifies capability. This work demonstrates that GenAI's promise lies not in individualized tutoring but in catalyzing collective intelligence through collaborative pedagogy that maintains \textit{human reasoning as primary}, fundamentally a choice about whether engineering education optimizes for efficiency or depth, shaping not just how students learn but what kind of engineers they become.

\section*{Acknowledgment}
The authors thank participating students and acknowledge the use of AI tools (ChatGPT, Claude, Grammarly) for writing, editing, and formatting support. All the content is verified, and the authors take full responsibility for accuracy and integrity.

\bibliographystyle{IEEEtran}
\bibliography{GCI}

\end{document}